\def\mb{M_{\rm B}}

\def\msol{{\rm M}_{\odot}}

\def\flm{f_{\rm 1,infx}}

\def\mblm{M_{\rm B}^{\rm (1,infx)}}

\def\mb{{M}_{\rm B}}
\def\rhoc{\rho_{\rm c}}

\documentclass{aa}
\usepackage{graphicx}
\begin{document}
%
%
%
\newcommand{\ddp}[2]{\frac{\partial #1}{\partial #2}}
\newcommand{\ddps}[2]{\frac{\partial^2 #1}{\partial #2 ^2}}

\title{Hyperon softening of the EOS of dense matter
and the spin evolution of isolated neutron
stars}
\author{J.L. Zdunik\inst{1}
\and
P. Haensel\inst{1,2}
 \and
E. Gourgoulhon\inst{2}
\and
M. Bejger\inst{1}
}
\institute{N. Copernicus Astronomical Center, Polish
           Academy of Sciences, Bartycka 18, PL-00-716 Warszawa, Poland
{\em jlz@camk.edu.pl}}
\institute{N. Copernicus Astronomical
Center, Polish
           Academy of Sciences, Bartycka 18, PL-00-716 Warszawa, Poland
\and LUTH, UMR 8102 du CNRS, Observatoire de Paris, F-92195 Meudon Cedex, France\\
{\tt jlz@camk.edu.pl,  haensel@camk.edu.pl,
  Eric.Gourgoulhon@obspm.fr, bejger@camk.edu.pl}}
\offprints{J.L. Zdunik}
\date{Received  / Accepted }
\abstract{The effect of the hyperon softening of the equation of state (EOS)
of dense matter on the spin evolution of isolated neutron stars is studied
for a broad set of hyperonic EOSs. We use a multidomain 2-D code
based on a spectral method, and show how important the precision
of solving the equations of stationary motion is for the stability analysis.
For some EOSs, the hyperon softening leads to spin-up by the angular momentum
loss, in the form of the back bending phenomenon, for a rather broad range
of stellar baryon mass. We show that large segments of the evolutionary
tracks exhibiting  the back bending behaviour in the moment-of-inertia -
rotation-frequency plane, are actually unstable and therefore not
astrophysically relevant. We show also that during the
spin-up - angular-momentum-loss
epoch, an isolated neutron star (e.g., a radio pulsar) can lose a
sizable part of its initial angular momentum without significantly
changing its
rotation period.  We propose also simple arguments and criteria
allowing  to connect the presence of a back bending epoch  with the
mass-radius relations  and the stiffness and/or softness of the nucleon and
hyperon EOSs of the neutron star core.
 }
\titlerunning{Hyperons and back bending in neutron stars}
\maketitle


\section{Introduction}
\label{sect:introduction}
Some theories of dense matter predict a softening of the equation of state (EOS) at
densities exceeding normal nuclear density $\rho_0=2.7\times 10^{14}~{\rm g~cm^{-3}}$,
implied by a phase transition to a new ``exotic'' phase. Several ``exotic''
high-density phases were proposed in the past, e.g., deconfined quark plasma, and
pion-condensed or kaon-condensed  hadronic matter. The softening of the EOS could be
due to a transition into a pure ``exotic'' phase, or to a mixture of an ``exotic''
phase with normal phase of dense matter. It has been suggested that the softening of
the EOS due to a phase transition could lead to characteristic back bending phenomenon
in the timing of spinning-down pulsars (Glendenning et al. 1997) or produce
characteristic period clustering in spinning-up neutron stars in low-mass X-ray binaries
(Glendenning 2001, Glendenning and Weber 2002). An interesting conclusion
of these papers was that both back bending in spinning down pulsar and spin clustering
in accreting millisecond pulsars stars is an {\it evidence} of a phase transition
taking place at the center of a spinning-down pulsar or a
spinning-up accreting neutron star.

As we show in the present paper, using several EOS of dense matter and very precise
code for the calculation of the rotating stellar configurations,  observation of the
back bending in the timing behavior of isolated pulsars, or of the
period  clustering,  is not an  unambiguous  evidence for an ``exotic''
phase in dense matter. Namely, these phenomena can also be implied  by the presence of
hyperons in dense matter, a feature which is in no way ``exotic'' and which was predicted
more than forty years ago (Cameron 1959, Salpeter 1960, Ambartsumyan and Saakyan
1960). A possibility of ``spin-up by the angular momentum
loss'' for a normal sequence (baryon mass smaller than the maximum baryon mass of
static configurations) of spinning-down neutron stars with hyperonic cores was
previously noted by Balberg et al. (1999).
However, as we will demonstrate in the
present paper, a complete study of back bending requires
a very precise ``exact'' code
for calculating stationary configurations
of rotating stars and a simultaneous careful
checking of secular stability of these configurations. Such
 conditions  were typically not satisfied in previous calculations.

In previous works the  back bending phenomenon has been considered as a feature of
the $I(\Omega)$ dependence, where $I$ is the moment of inertia of the star and
$\Omega$ is the angular frequency of rotation (\cite{gpw97}, \cite{cyz02} and
\cite{ss02}). In the present  paper we clarify some statements about the
back bending for rotating neutron stars, and we formulate some rules which are useful
for searching for the back bending in rotational stellar sequences. As we demonstrate
using high-precision evolutionary sequences, the whole analysis should be
performed using different pair of variables: total stellar angular momentum $J$ versus
$\Omega$ instead of $I(\Omega)$. High precision is particularly important because it
is needed to reliably check the secular stability of rotating configurations; only the
stable ones are interesting and observationally relevant.

The paper is organized in the following way. Softening of the EOS
due to the presence of hyperons is discussed in Sect.\
\ref{sect:EOS}. The method allowing for a  high precision of
the 2-D calculations of the equilibrium configurations of rotating
neutron stars is briefly described in Sect.\ \ref{sect:NumCalc}. Different
formulations of the stability criteria for rotating configurations
are briefly summarized in Sect.\ \ref{sect:I.J.stability}. In
Sect.\ \ref{sect:MR.minima} we propose a method of checking for
the occurrence of the back bending phenomenon by inspecting the
baryon-mass -- equatorial-radius relations at fixed values of
rotation frequencies. We apply this method  to several  EOSs with a hyperon
softening. The interplay between the back bending and stability is
discussed in Sect.\ \ref{sect:Jf.BB.stab}, where we study
neutron-star evolution tracks in the angular-momentum --
rotation-frequency plane. In Sect.\ \ref{sect:BB.EOS} we study the
dependence of the back bending phenomenon on the EOS of the
hyperonic matter. Final Sect.\ \ref{sect:DiscConcl} contains
discussion of our results, including their possible observational
aspects, and ends with  concluding remarks.

\section{Equation of state with hyperons}
\label{sect:EOS}
\begin{figure}
\resizebox{\hsize}{!}{\includegraphics{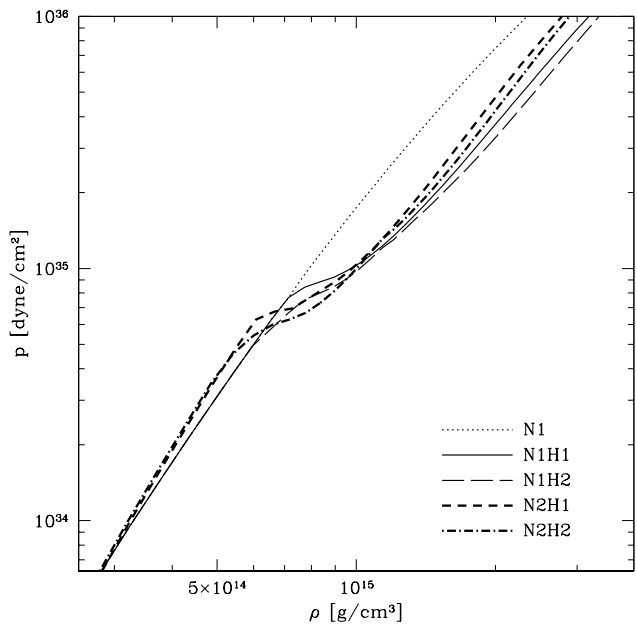}}
\caption{Equations of state of hyperonic matter calculated by
Balberg \& Gal (1997) and used in this paper. Our notation is a close
analogue of that  introduced by Balberg et al. (1999). Our labels
N1, N1H1, and N1H2 are their EoS1 N, EoS1 N$\Lambda\Xi$,
and EoS1 N$\Lambda\Sigma\Xi$, respectively. Our N2H1 and N2H2 are
their EoS2 N$\Lambda\Xi$ and EoS2 N$\Lambda\Sigma\Xi$,
respectively. For further explanations see the text.}
\label{fig:EOS}
\end{figure}
Possible presence of hyperons in dense neutron-star matter is mainly
the consequence of  the Pauli principle for neutrons and electrons, which
at a sufficiently high baryon density can make a replacement of high-energy
neutrons and electrons by more massive but slower hyperons energetically
favorable. Hyperon species $H$ is then present above a certain
threshold $\rho_{H}$,
which is determined by a condition involving the change of energy of dense
matter,  due to addition of a single hyperon, at a fixed pressure, $\mu_{H}^0$.
The threshold density for $H$ is the lowest density at which the equality
\begin{equation}
\mu^0_{H}=\mu_n -q_{H}\mu_e~,
\label{eq:rho.H}
\end{equation}
is satisfied. Here, $\mu_n$ and $\mu_e$ are the chemical potentials of neutrons and
electrons (which include rest energies), and $q_{H}$ is the hyperon charge in
units of $e$. The threshold condition given by Eq.\ (\ref{eq:rho.H}) shows that
because of the high value of $\mu_e$,  hyperons with negative charge are strongly
preferred over the positively charged ones. It also explains, why $\Sigma^-$, and not
the lowest mass hyperon, $\Lambda^0$,  is usually the first hyperon to appear in dense
matter.  It should be stressed, however, that the hyperon-nucleon interactions
which contribute significantly to $\mu_{H}^0$, are poorly known, and this
implies uncertainty not only in the values of $\rho_{H}$  but also
in the order of the hyperon  appearance. Let us notice that for some models
of dense matter hyperons do not appear in neutron stars at all, because
the lowest $\rho_{H}$ is larger than the maximum density in neutron stars
(see, e.g., Pandharipande \& Garde 1972).

 The hyperon softening is particularly well visualized by the behavior of
the adiabatic index
\begin{equation}
\Gamma={n_{\rm b}\over P}{{\rm d}P\over {\rm d}n_{\rm b}}~.
\label{eq:Gamma}
\end{equation}
At each threshold density $\rho=\rho_{H}$
the function $\Gamma(\rho)$ suffers a
drop. For some models, typical values of $\Gamma\simeq 2-3$ characteristic
of the nucleonic matter at $\rho\sim (2-4)\rho_0$ can drop even down
to $\Gamma\sim 1$ (see, e.g., Balberg et al. 1999). For other models
the drop is not so dramatic but still sizable (see, e.g., Haensel et al.
2002). The softening of the EOS by hyperonization of matter can be also
visualized by comparing the values of the maximum allowable mass for
non-rotating neutron stars, $M_{\rm max}$,  for the EOS without hyperons,
referred hereafter as the N EOS, and those involving nucleons and
hyperons (NH). Typically, allowing for the presence of hyperons
lowers the value of $M_{\rm max}$ by $(0.3-0.6)~M_\odot$ (see, e.g.,
Haensel 2003). The presence of hyperons leads to a very characteristic
flattening of the mass-radius and mass-central density plot for neutron
star, with a knee taking place just after the threshold for the first hyperon
(Glendenning 1985, Balberg et al. 1999).
This feature will be important in the context of the back bending phenomenon
in rotating neutron stars.

 In present paper we use the EOSs  calculated by
Balberg and Gal (1997). These EOSs are presented in Fig.\ \ref{fig:EOS}.
They are based on  phenomenological  effective
baryonic matter energy functionals, resulting from
effective baryon-baryon interactions whose parameters are
adjusted to reproduce the empirical properties of nuclear
matter as well as the basic experimental features of the
hyperonic interactions. Two models of nucleon matter,
based on an effective nucleon-nucleon (NN) interaction,
constructed by Balberg and Gal (1997) lead to the EOSs
N1 and N2 of dense matter without hyperons (they
correspond to  their models EoS1 N and EoS2 N, in the notation of
Balberg et al. 1999). The model N1 corresponds to the
incompressibility of nuclear matter at saturation $K_0=240$ MeV
(i.e., is close to the experimental value of this parameter).
The model N2 gives $K_0=320$ MeV, and leads to a significantly
stiffer EOS than the N1 one. Balberg \& Gal constructed also
 two models representing effective
 nucleon-hyperon  (NH)  and hyperon-hyperon (HH)  interactions in baryonic
matter. We will denote them by H1 and H2; they correspond
to the $\Lambda\Xi$ and $\Lambda\Sigma\Xi$ effective hyperon
interaction models of Balberg \& Gal (1997) in the notation of
Balberg et al. (1999). Model H2 leads to a stronger softening
of the EOS due to the presence of hyperons
than the H1 one. The EOS of baryonic
matter are obtained by combining NN, NH, anf HH effective
interactions. In this way one obtains four EOSs of baryonic
matter with hyperons. We will denote them by N1H1, N1H2,
N2H1, and N2H2; they correspond to  EoS1 N$\Lambda\Xi$,
 EoS1 N$\Lambda\Sigma\Xi$, EoS2 N$\Lambda\Xi$ and
EoS2 N$\Lambda\Sigma\Xi$ in the notation of Balberg et al. (1999),
respectively. For the sake of completeness, we considered also five EOSs
of Glendenning (1985); see Sects. \ref{sect:BB.EOS} and \ref{sect:DiscConcl}.

\section{Numerical calculations}
\label{sect:NumCalc}

The neutron star models have been computed in full general relativity
by solving the Einstein equations for stationary axi-symmetric spacetime
(see e.g. \cite{BonazGSM93} or \cite{GourgHLPBM99} for the complete
set of partial differential equations to be integrated).
The numerical computations have been performed via
the {\tt Lorene/Codes/Rot\_star/rotstar} code
from {\sc Lorene} ({\tt http://www.lorene.obspm.fr/}).
This C++ code implements a multi-domain spectral method
introduced in \cite{BonazGM98}.  A description of
the code can be found \cite{GourgHLPBM99}. For the
purpose of the present work, we have employed two
domains to describe the neutron star interior, making
use of the adaptive coordinates to set the boundary
between the innermost domain and the outer one
at the transition surface to hyperon matter.
In this way, the density field is smooth in each domain
and the spectral method results in a high accuracy.
This accuracy has been checked by evaluating the
GRV2 and GRV3 virial error indicators
(see \cite{NozawSGE98}), which showed a relative
error lower than $\sim 10^{-5}$.

The physical parameters resulting from the equation of state
(pressure, energy density, number density) are obtained by the
Hermite interpolation (\cite{Swesty96, NozawSGE98}).
The important feature of this approach is
automatic fulfillment of the first law of thermodynamics (the
Gibbs-Duhem relation) (see \cite{NozawSGE98}).

\section{Back bending phenomenon, $I$ and  $J$ vs. $\Omega$,
 and stability of rotating configurations}
\label{sect:I.J.stability}
The term  back bending comes from  nuclear physics (see,~e.g.,  Ring \& Shuck 1980).
Nuclei can be excited by a projectile
to a state of a rapid rotation corresponding to a large angular momentum
quantum number ${\cal J}$ and excitation energy $E({\cal J})$.
The nuclear angular momentum is measured in the units of $\hbar$.
The eigenvalues of the
operator of the square of the angular momentum are ${\cal J}({\cal J}+1)$.
For ${\cal J}\gg 1$ one can approximate this eigenvalue  by a ``classical value''
${\cal J}^2$. Within the quasiclassical approximation (in which
${\cal J}$ can be treated as a continuous quantity) one  can
phenomenologically  define an ``angular frequency'' by $\omega =
{\rm d}E/{\rm d}{\cal J}$. The nuclear moment of inertia ${\cal I}$ is
then found by fitting the
rotational $E({\cal J})$ spectra. In the quasi-classical approach,
${\cal I}$ is a function of $\omega$, which can be represented by a
curve in the ${\cal I}-\omega$ plane:
along this curve, ${\cal J}$ increases.
In the standard case there is a one to one
correspondence between ${\cal I}$ and $\omega$, and  both ${\cal I}(\omega)$
and $\omega({\cal I})$ are increasing functions of their arguments. However,
for some nuclei  (e.g., $^{158}{\rm Er}$ see Ring \& Schuck 1980)
$\omega$ reaches a maximum at some value of ${\cal J}={\cal J}_1$ and then
 decreases (back bends) to reach minimum at some larger value of ${\cal
J}={\cal J}_2$. At ${\cal J}={\cal J}_1$ and ${\cal J}={\cal J}_2$ the derivatives
${\rm d}\omega/{\rm d}{\cal J}$ and ${\rm d}\omega/{\rm d}{\cal I}$ vanish, and for
${\cal J}_1<{\cal J}<{\cal J}_2$ the curve $\omega({\cal I})$
is ``back bending''  which corresponds to $\omega$ which is decreasing with
increasing ${\cal I}$ (see Fig. 3.4 in Ring \& Schuck 1980).

Neutron star can be treated as a huge atomic nucleus. However, such a star is a
macroscopic  classical object containing some $\sim 10^{57}$ baryons,
to be compared with at most $\sim 200-250$
nucleons in rapidly rotating nuclei. In nuclei, all high--angular-momentum states
are the excited ones and therefore unstable. In the case of rotating neutron stars
unstable states are not interesting, their lifetime being too short to observe them.
Therefore
in the case of neutron star we have to check whether a given state of stationary
rotation is stable, because instability would make it astrophysically irrelevant.
In this context, we  find it convenient to
discuss the back bending phenomenon by studying the dependence of  the total angular
momentum of the star $J$  versus frequency of rotation $f=\omega/2\pi$.
We differ in this choice from the previous
work, in which the dependence $I(f)$ was studied.
The reason of this choice is the following:
 $J$ is a well defined quantity describing instantaneous state of
rotating relativistic star and  the evolution of rotating star can be easily
calculated under some assumptions about the change of $J$.
In what follows, we restrict ourselves to evolutionary tracks of
isolated neutron stars for which baryon mass $M_{\rm B}=const$.

The moment of inertia $I$ is  usually defined as $J/\Omega$
(see, e.g.,  Stergioulas 2003). This is the definition used in the previous
papers on the back bending phenomenon in rotating neutron stars.
However, such  $I$ {\it does not} describe the response  of the star to the
change of $J$ or $\Omega$ and therefore is not useful for checking the
stability of rotating configuration. In order to obtain, e.g., the spin down
of a  star due to the angular momentum decrease ${\rm d}J$
one should have defined   $\widetilde{I}\equiv
{\rm d}J/{\rm d}\Omega$. Only in the slow rotation limit, where only
terms linear in $\Omega$ are conserved, both definitions of the moment of
inertia coincide.
\begin{figure}
\resizebox{\hsize}{!}{\includegraphics[]{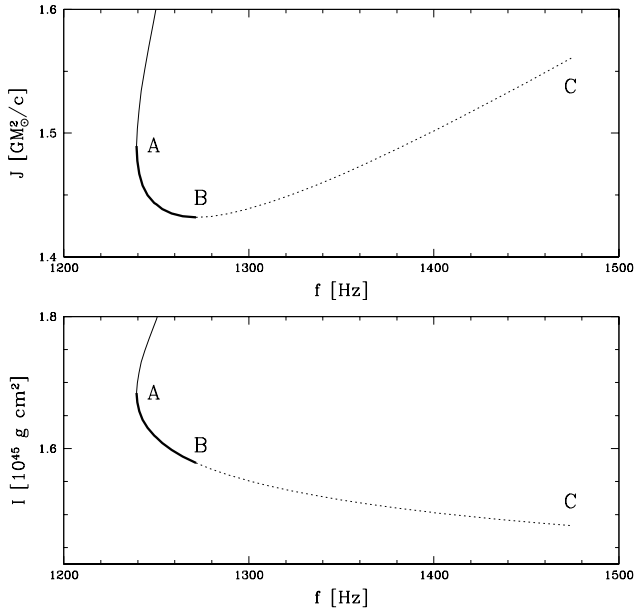}}
\caption{The difference between the back bending curves $J(f)$ and
$I(f)$. The spin-evolution track is calculated for  $M_{\rm
B}=2.15~{\rm M}_\odot$ and N2H1 EOS. Upper  panel presents the
dependence $J(f)$. The thicker segment AB represents stable back
bending evolution, while dotted segment to the right of the
minimum at $B$ consists of configurations unstable with respect to
axisymmetric perturbation. The lower panel represents the $I(f)$
spin evolution track, considered in the previous work; the back
bending behavior occurs apparently along the whole AC branch.
Actually, the BC segment of the  $I(f)$ is astrophysically
irrelevant, because configurations to the right of B are unstable.
}
 \label{fig:unstable}
\end{figure}
%

Total stellar angular momentum $J$ is not only
a quantity with a strict physical meaning
in general relativity. It  also allows us to study
the stability of rotating configurations with respect to axially
symmetric perturbations. The point of the
change in stability  within a family
of rotating  configurations
(from stable to unstable or {\it vice versa}).
corresponds to the extremum of $M$ or $\mb$ at fixed $J$ (\cite{Fried1988}):
\begin{equation}
\left({\partial M\over \partial x}\right)_{J={const}} =0~,
~~~~~~~~~\left({\partial\mb \over \partial x}\right)_{J={const}} =0~,
\label{eq:M.MB.J.stab}
\end{equation}
where $x$ is  the first  of the two  parameters which parametrize (label)
 stationary rotating stellar configurations (the second one
 being $J$),  for example
  $x=\rho_{\rm c}$ or  $x=P_{\rm c}$.
Equivalently,  the onset of instability can be determined  by the
condition
\begin{equation}
\left({\partial J\over \partial x}\right)_{\mb=const} =0~,
~~~~~~~~~\left({\partial J \over \partial x}\right)_{M={const}} =0~.
\label{eq:J.MB.stab}
\end{equation}

\begin{figure}
\resizebox{\hsize}{!}{\includegraphics[angle=-90]{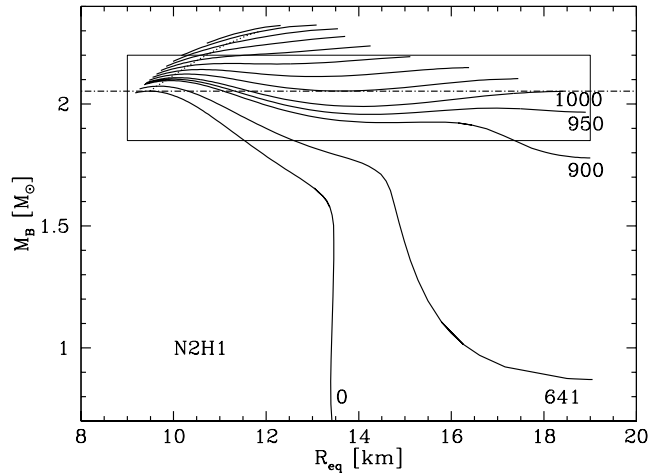}}
\caption{Total baryon mass $M_{\rm B}$  vs circumferential equatorial radius
$R_{\rm eq}$ for stationary configurations rotating at a  fixed
frequency, for the N2H1  EOS. The curves are labeled by the rotational
frequency $f=\Omega/2\pi$ [Hz].  The dotted line corresponds to the
onset of instability with respect to axi-symmetric
(quasi-radial) perturbations. All curves terminate on the
large-radius side at the mass-shedding (Keplerian) configurations.
The maximum baryon mass of non-rotating stars, $M_{\rm B,max}^{\rm
stat}$, is marked by a dash-dotted line.}
\label{fig:mbrn2h1}
\end{figure}
%
\begin{figure}
\resizebox{\hsize}{!}{\includegraphics[angle=-90]{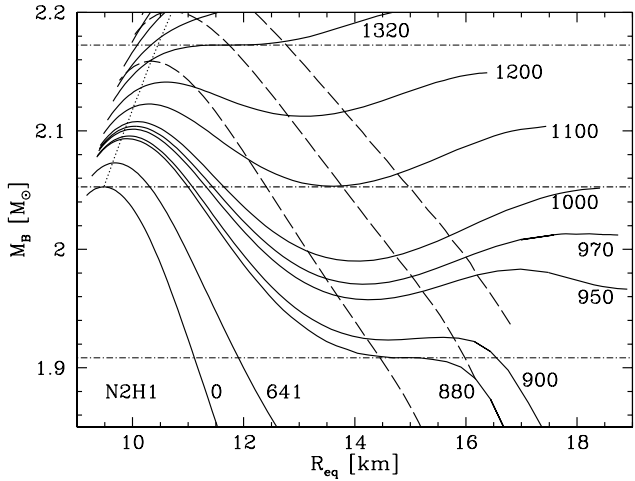}}
\caption{The enlarged region of the back-bending phenomenon
(corresponding to the box in Fig. \ref{fig:mbrn2h1}). For fixed rotational frequency
$f_{\rm 1,infx}\simeq 880$~Hz and
$f_{\rm 2,infx}\simeq 1320$~Hz the $\mb(R_{\rm eq})$ dependence
has an inflexion point (corresponding to the masses $M_{\rm
B}^{\rm (1,infx)}=1.91~\msol$ and $M_{\rm B}^{\rm (2,infx)}=2.17~\msol$,
respectively) resulting in a  region where the curve is nearly flat.
For $f \in [f_{\rm 1,infx},f_{\rm 2,infx}]$ there exists  a  local
minimum of $\mb$. Dashed curves correspond to a  fixed total
angular momentum $J$. The configurations close to the local maxima
are obviously stable  (for a fixed $J$, $\mb$ is monotonic). The
evolution of an  isolated star which is losing its angular
momentum is represented by the motion along a horizontal line from
right to the left (decreasing $J$). The loss of $J$ in the back
bending regime is associated  with a spin up of the star.}
 \label{fig:mbrn2h1x}
\end{figure}
\begin{figure}
\resizebox{\hsize}{!}{\includegraphics[angle=-90]{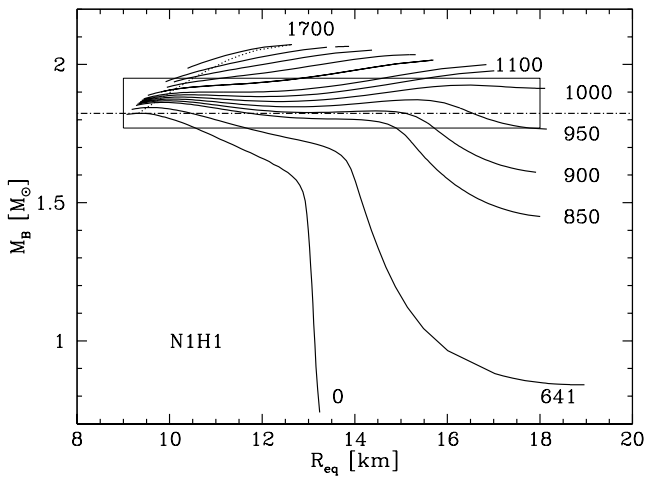}}
\caption{Same as in Fig.\ \ref{fig:mbrn2h1} but  for the N1H1  EOS.
}
 \label{fig:mbrn1h1}
\end{figure}
%
\begin{figure}
\resizebox{\hsize}{!}{\includegraphics[angle=-90]{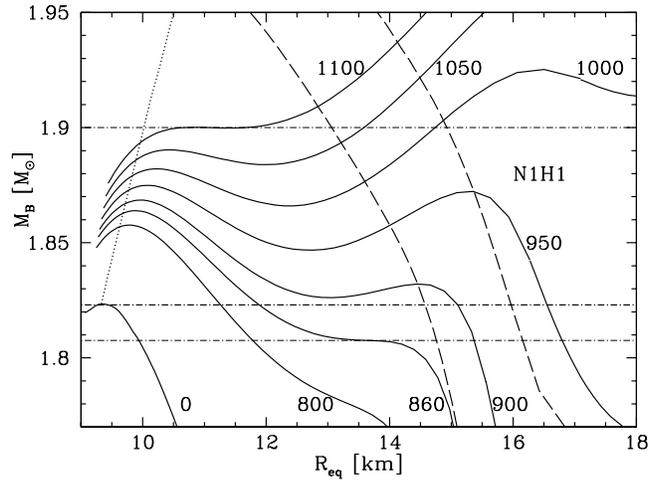}}
\caption{The enlarged region of the back-bending phenomenon
(corresponding to the box in Fig. \ref{fig:mbrn1h1}).}
 \label{fig:mbrn1h1x}
\end{figure}

The crucial  difference between the dependencies $J(f)$ and $I(f)$
in  the back-bending region is presented in Fig.\
\ref{fig:unstable}; it will be discussed in more detail in
Sect.\ \ref{sect:Jf.BB.stab}. The apparently ``back bending branch'' in lower
panel, as defined in the previous work (Glendenning et al. 1997),  consists mostly of
configurations unstable with respect to axial perturbations. The BC segment is
astrophysically irrelevant, and the B point correspond to an unstable termination
of the back bending branch (see Sect.\ \ref{sect:BB.ut.st}).
\section{Softening of the EOS and the baryon-mass--equatorial-radius
 relation at fixed rotation frequency }
\label{sect:MR.minima}
Rotation modifies  the relation between the baryon mass ($M_{\rm B}$) and the
circumferential equatorial radius ($R_{\rm eq}$) for equilibrium stellar
configurations. The baryon mass $M_{\rm B}$ plays a special role,
 because it remains constant during the evolution of
 solitary  pulsars. In the present section we point out specific features
of the $M_{\rm B}(R_{\rm eq})$ curves, which signal
 the presence of the back bending (BB) phenomenon in rotating neutron stars.
 Our calculations were performed for all the equations of state with
hyperons presented in \cite{blc99}.
However, we present detailed  results (figures) only for those EOSs for which the BB
phenomenon is strongly pronounced.

In Fig.\ \ref{fig:mbrn2h1} we show the $M_{\rm B}(R_{\rm eq})$
curves for stars rotating  at a fixed rotation frequency
$f=\Omega/2\pi$, calculated for the N2H1 EOS.
 An enlarged view
of a particularly interesting rectangular region of the $M_{\rm
B}-R_{\rm eq}$ plane is shown in Fig.\ \ref{fig:mbrn2h1x}.

\subsection { Signature of BB: minimum of the baryon mass at
fixed frequency}
As we will show, the BB  phenomenon is strictly connected with the
existence of a local minimum of $\mb$ in the $M_{\rm B}(R_{\rm eq})$ plot at a
fixed $f$. The softening of the EOS  due the hyperonization
leads to the flattening of the
$M_{\rm B}(R_{\rm eq})$ and $M_{\rm B}(\rho_{\rm c})$ curves (for the
case of non-rotating stars, see Balberg et al. 1999).
This effect of flattening grows with increasing  rotational frequency and for
 $f>\flm $ rotation may  even produce  a  local minimum of
 $\mb(R_{\rm eq})$ and $M(R_{\rm eq})$.
 The curve $M_{\rm B}(R_{\rm eq})$ [or
 $M_{\rm B}(\rho_{\rm c})$] at $f=\flm $ has a very specific property.
Namely, for this curve  first and second derivatives of $\mb$ with
respect to the central density  $\rho_{\rm c}$ vanish  at some
$\rho_{\rm c}= \rho_{\rm c, infx}$, i.e.,  the curve  has there a
point of inflexion which corresponds to baryon mass $\mblm$.
Our numerical calculations for the N2H1 and
N1H1 EOS give $\flm \simeq 880$~Hz and $\flm \simeq
860$~Hz respectively (see Figs.\ \ref{fig:mbrn2h1x} and \ref{fig:mbrn1h1x}). For
$f>\flm $ the curve $M_{\rm B}(R_{\rm eq})$, $M_{\rm B}(\rho_{\rm
c})$, \ldots, exhibit a local minimum. We find that the  presence
of this local minimum is an indication that for $M_{\rm B}>
\mblm$ the rotational evolution  of neutron star with $\dot{J}<0$
exhibits a BB phenomenon in the vicinity of $f\simeq \flm $. This
is clearly seen in Figs.\ \ref{fig:mbrn2h1x} and \ref{fig:mbrn1h1x},
where we show an enlargement of the vicinity of the inflection
point, together with $J=const.$ lines. As we see, in this case
there exist a range of $M_{\rm B}$ where  the decrease of $J$ leads to the
increase of the angular frequency which is exactly equivalent to
back bending. The fragment of the curve for which $M$ decreases as
a function of $\rho_{\rm c}$ does not necessarily correspond to
the instability region - the decrease of $M_{\rm B}$ at a fixed
$f$ does not imply the decrease at a fixed $J$. It is only the
latter condition which indicates the instability with respect to
small axi-symmetric perturbations.
\footnote[1]{It may be shown that the  configurations which
realize the extrema of the
$M_{\rm B}$ and $M$ at fixed $J$ coincide, see Harrison et al. 1965
for the static case, and Friedman et al. (1988)  for uniformly rotating
configurations.}

In Figs.\ \ref{fig:mbrn2h1x} and \ref{fig:mbrn1h1x}
 we draw also three horizontal lines
corresponding to  fixed values of the total baryon number. A
rotating star losing its angular momentum  moves along horizontal line
from the right to the left. The bottom lines correspond to the
mass  $M_{\rm B}=\mblm \simeq 1.91~\msol$ ($\simeq 1.81~\msol$)
for the N2H1 (N1H1) EOS,  at  which the
curves for $f=\flm  \simeq 880$~Hz ($860$~Hz) have a
point of inflexion.
The top horizontal line corresponds to the different  situation in which
the curve $M_{\rm B}(R_{\rm eq})$ has an inflexion point at a
higher frequency, namely at $f=1320$  Hz  and
1100 Hz, for the N2H1 and N1H1 EOSs, respectively.
For baryon masses larger than the mass at this inflexion point,
the angular momentum loss does not lead to
the decrease of angular frequency before the onset of instability
is reached, the star is all the time accelerating.
The value of this limiting masses are  $\mb\simeq
2.17~\msol$ and $ 1.90~\msol$, respectively.
The intermediate horizontal line corresponds to the
maximum mass of the non-rotating stars $M_{\rm B}=M_{\rm
B,max}^{\rm stat}=2.05~\msol$ and $1.824~\msol$.

\subsection{Acceleration or slowing down close to the Keplerian
limit}
The interesting difference between the cases N2H1 and N1H1 concerns
the behavior of the rotating star as it starts losing $J$  at the
Keplerian frequency $f_{\rm K}$. The question is whether
the star is then slowing down or spinning up.
The actual behavior  can  be  easily deduced  from the shapes of the
$\mb(R_{\rm eq})$ [or $\mb(\rhoc)$] curves for a fixed frequency close to the Keplerian limit.
If $\mb$ is increasing as we move in the $M_{\rm B}-R_{\rm eq}$ plane
away from the Keplerian configuration [$({\rm d}\mb/{\rm d}\rhoc)_{\rm K}>0$],
the star is slowing down as it loses the angular momentum. This effect results in the
S-shape of the $J(f)$ dependence in the back-bending case.
Otherwise [$({\rm d}\mb/{\rm d}\rhoc)_{\rm K}<0$]
the isolated star is spinning up when evolving from the Keplerian configuration
with $\dot{J}<0$.
The limiting case corresponds to the condition
$({\rm d}\mb/{\rm d}\rhoc)_{\rm K}=0$ [or $({\rm d}\mb/{\rm d} R_{\rm eq})_{\rm
K}=0$]
and is represented  in Fig. \ \ref{fig:mbrn2h1x}
by the curve with $f=970$~Hz (the baryon mass at this point
$M_{\rm B}^{\rm (K,flat)}=2.01~\msol$). For the N1H1 EOS such an  effect appears for
supramassive stars ($f\simeq 1030$~Hz, $M_{\rm B}^{\rm (K,flat)}=1.96~\msol$,
see Fig.\ \ref{fig:mbrn1h1x})
%
\section{Back bending  and stability:
analysis in the $J-f$ plane}
\label{sect:Jf.BB.stab}
\begin{figure}
\resizebox{\hsize}{!}{\includegraphics[angle=0]{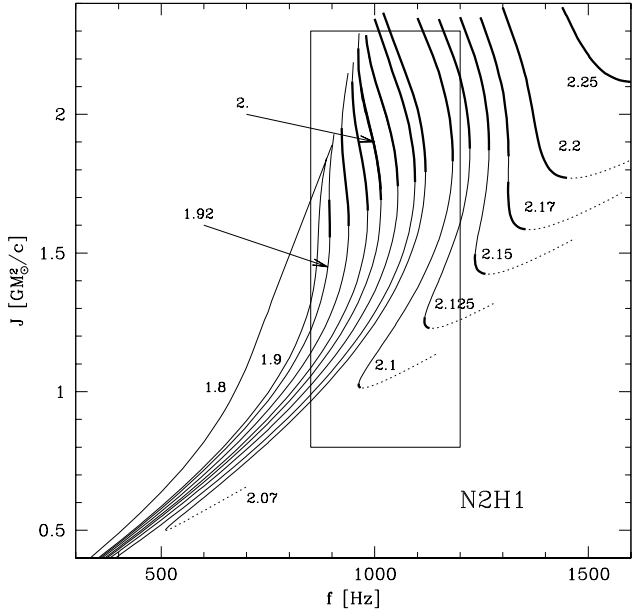}}
\caption{Angular momentum of the star $J$ versus rotation
frequency $f$, for the N2H1 EOS.  Each curve corresponds to a
fixed $M_{\rm B}$, whose value in ${\rm M}_\odot$ is displayed.
Along each curve, the central density increases downwards. The
dotted segments correspond to configurations which are unstable
with respect to axi-symmetric perturbations whereas the thick
lines correspond to spin-up by angular momentum loss. An enlarged
view of the rectangular region within which the back bending
occurs is shown in Fig. \ref{fig:jfn2h1x}.  }
 \label{fig:jfn2h1}
\end{figure}

The search for the  BB  phenomenon with simultaneous
testing  of the stability of rotating configurations  can be most
conveniently carried out by plotting, at a fixed $M_{\rm B}$,
 the stellar angular momentum $J$  versus rotation frequency $f$.
 Let us start with the N2H1 EOS where the BB is the most pronounced.
 Several  curves $J=J(f)$ at selected values of $M_{\rm B}$,
 calculated for this  EOS, are shown in   Fig.\ \ref{fig:jfn2h1}. These
 curves represent  the evolution  of  an isolated pulsar of a given baryon mass
 $M_{\rm B}$, as it  loses its angular momentum due to radiation of
electromagnetic waves. Along each curve, the central density
$\rho_{\rm c}$ increases monotonically when one moves downward.
 For stable configurations $J$ is a monotonic function
along this path. Any minimum  indicates the onset of the instability
with respect to axi-symmetric perturbations.
Putting it differently, for stable configurations  each value of
$J$ corresponds to one and only one value of $f$.
The BB manifests itself as a stable segment of the  $J(f)$ curve
with ${\rm d}J/{\rm d}f<0$.

It should be mentioned that there is an important difference
between the information one can get from the analysis of the
$J(f)$ and the usually used $I(f)$ curves. Although the  $I(f)$
can have segments with ${\rm d}I/{\rm d}f<0$ corresponding  to a
back bending in a nuclear physics sense (Glendenning et al. 1997),
the $I(f)$ dependence cannot tell us whether a seemingly ``back
bending branch'' contains configurations which are stable. One can
have a minimum in $J(f)$ on a ``back bending'' segment of $I(f)$
(see Fig.\ \ref{fig:unstable}); such a possibility was already
mentioned by Spyrou \& Stergioulas (2002).

The final fate of the rotating star as its angular momentum decreases depends
 on $M_{\rm B}$. If $\mb > M_{\rm B,max}^{\rm stat}$, the star is supramassive and
eventually collapses into a black hole.

 For the N1H1 EOS the BB phenomenon is less pronounced than for the N2H1 one.
 The spin-evolution tracks at $M_{\rm B}=const$ are  shown in Figs.\ \ref{fig:jfn1h1}
 [$J(f)$ curves]  and \ref{fig:jfn1h1x} [zoomed $J(f)$ curves in the BB region for
 normal configurations]. The zoomed fragment in Fig.\ \ref{fig:jfn1h1x} shows
 how narrow, compared with the N2H1 case, is the range of baryon masses for
 normal configurations where BB with stable termination occurs.

Using the method developed in Sect.\ \ref{sect:MR.minima}, one can readily
formulate the following criterion.
If $\mblm >M_{\rm B,max}^{\rm stat}$ then the back bending is possible
only for supramassive configurations. In  the opposite case, the back bending
can occur also for rotating stars with $M_{\rm B}<M_{\rm B,max}^{\rm stat}$.

\begin{figure}
\resizebox{\hsize}{!}{\includegraphics[angle=0]{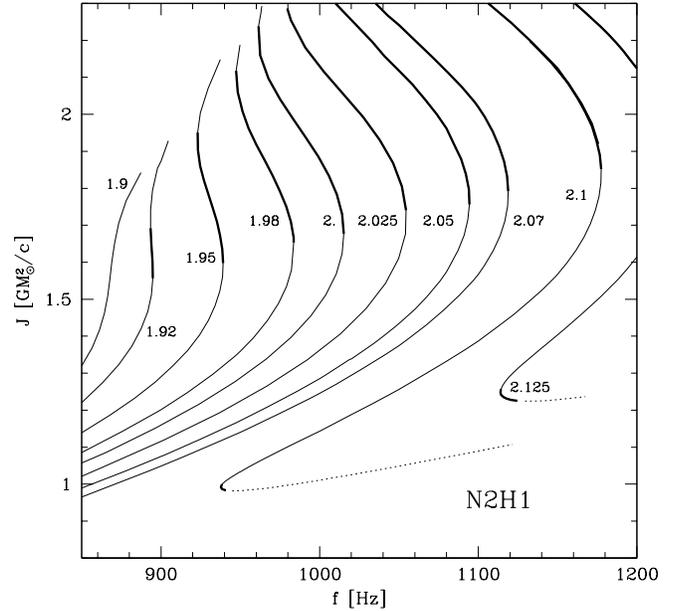}}
\caption{ Enlargement of Fig. \ref{fig:jfn2h1}. }
 \label{fig:jfn2h1x}
\end{figure}
\begin{figure}
\resizebox{\hsize}{!}{\includegraphics[angle=0]{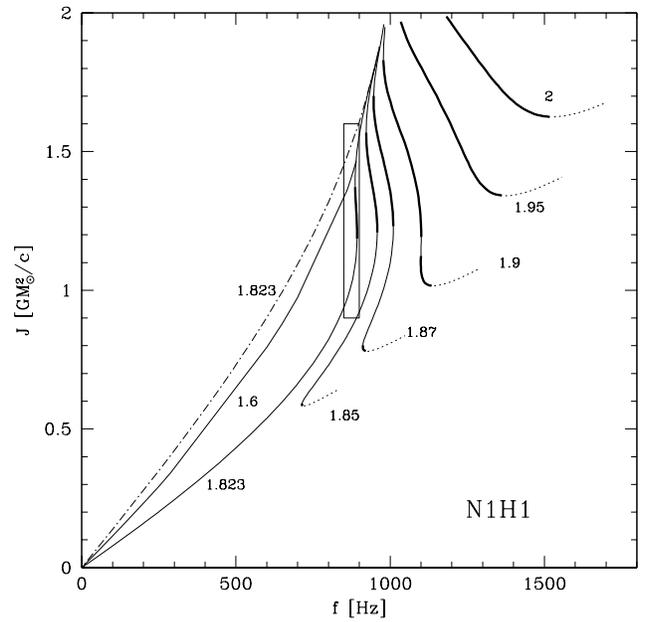}}
\caption{ Same as Fig \ref{fig:jfn2h1} but for N1H1 EOS. The
dash-dotted line is the $J(f)$ curve for the N1 EOS (i.e., not
allowing for the presence of the hyperons). }
 \label{fig:jfn1h1}
\end{figure}

\begin{figure}
\resizebox{\hsize}{!}{\includegraphics[angle=0]{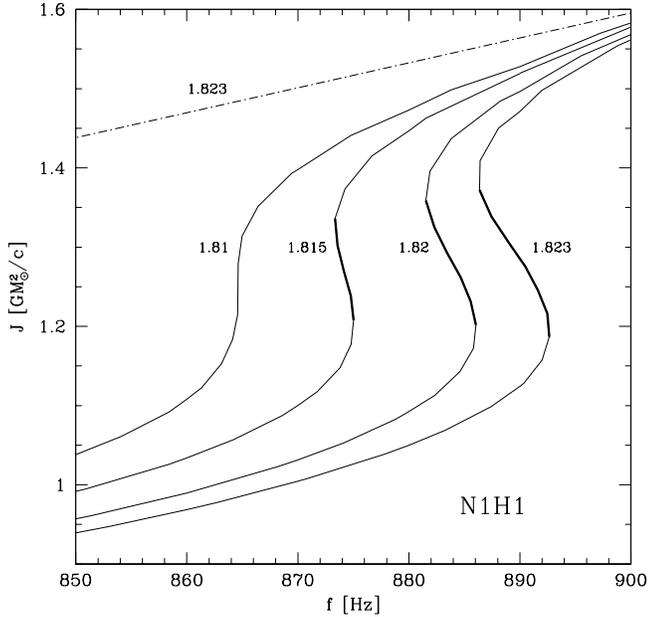}}
\caption{Angular momentum of the star as a function of frequency
for the N1H1 EOS, in the region where the back-bending phenomenon
with stable termination (${\rm BB_{st}}$, see the text) occurs.
Each curve corresponds to a fixed value of $M_{\rm B}$ (in ${\rm
M}_\odot$). Along each curve, the central density increases
downwards. Segments along which the angular momentum loss is
associated with a spin up are indicated by a thick solid line. All
configuration on the S-shaped thick segments are stable. The
dash-dotted line describes a $J(f)$ trajectory for the EOS of
matter in which hyperons are not allowed (N1 EOS, see the text).
This curve illustrates the dramatic effect of the hyperon
softening of the EOS on the pulsar spin-evolution.}
 \label{fig:jfn1h1x}
\end{figure}
\begin{figure}
\resizebox{\hsize}{!}{\includegraphics[angle=0]{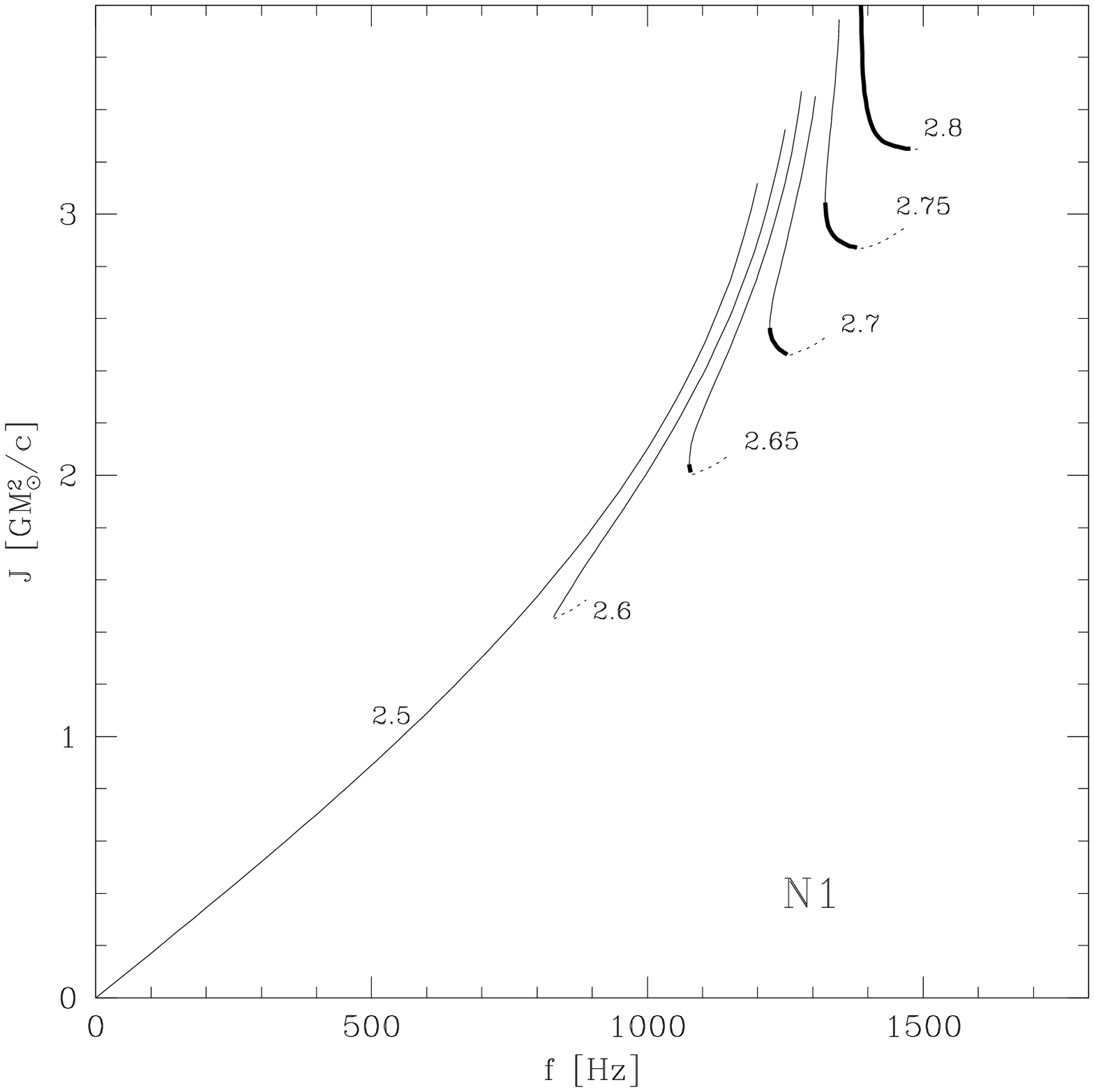}}
\caption{Angular momentum of a star with fixed  baryon mass
$M_{\rm B}$ as a function of frequency for the N1 EOS. Segments
constraining unstable configurations are dotted, those
corresponding to the spin-up by the angular momentum loss are
indicated by a thick solid line. }
 \label{fig:jfn1}
\end{figure}

\section{Two types of back bending}
\label{sect:BB.ut.st}
A look at Fig.\ \ref{fig:jfn2h1} shows that there are actually two
types of BB for rotating neutron stars. For isolated neutron star
$\dot{J}<0$ and therefore on a stable $J(f)$ branch $J_2<J_1$
implies $t_2=t(J_2)>t_1=t(J_1)$. Assume then that a BB epoch
starts at $t_1$ and terminates at $t_2$. For the first type of BB,
occurring for the baryon mass $2.25~{\rm M}_\odot$ and $2.2~{\rm
M}_\odot$ for the N2H1 EOS (Fig.\ \ref{fig:jfn2h1}), and for
$2~{\rm M}_\odot$ and $1.95~{\rm M}_\odot$ for the N1H1 EOS Fig.\
\ref{fig:jfn1h1}),  the point at $t_2$ corresponds to an unstable
rotating configuration. Such a BB  with an {\bf u}nstable {\bf
t}ermination will be denoted as $\rm BB_{ut}$. This type of BB
occurs above some value of $M_{\rm B}$ equal to $M_{\rm B}^{\rm
(2,infx)}$. An isolated neutron star enters the BB regime but it
will not  reappear in the normal spin-down regime  with ${\rm d}
J/{\rm d}f>0$. However, the situation changes as decreasing
$M_{\rm B}$ crosses the value $M_{\rm B}^{\rm (2,infx)}$. Then the
BB segment splits into two segments. The segment with higher $J$
(corresponding to an earlier epoch) terminates at a stable
configuration, and will be denoted by ${\rm BB_{\rm st}}$. Then
follows an epoch of normal spin down with ${\rm d}J/{\rm d}f>0$,
followed by a ${\rm BB_{\rm ut}}$ terminated by an instability.
With decreasing $M_{\rm B}$, the ${\rm BB}_{\rm ut}$ epoch becomes
shorter and shorter, and becomes infinitesimally small as one
reaches the maximum mass of nonrotating configuration $M_{\rm
B,max}^{\rm stat}$. For baryon masses larger than $M_{\rm B}^{\rm
(1,infx)}$ we are dealing with a ``BB episode'' in an otherwise
normal neutron star rotational evolution.

The case of  ${\rm BB}_{\rm ut}$ can be quite easily discussed on
the basis of the $\mb(R_{\rm eq})$ dependence (for example Fig.\
\ref{fig:mbrn2h1x}). The ${\rm BB}_{\rm ut}$ phenomenon
corresponds to a specific location of the instability line (dotted line
in Fig.\ \ref{fig:mbrn2h1x}) with respect to the maxima of the
functions $\mb(R_{\rm eq})$ at fixed $f$ (if they exist).
Because these maxima are to the right
of the instability line (i.e., in the stable region) ${\rm BB}_{\rm ut}$
always appears. In the other words,  a  star with fixed $\mb$,
 approaching the  instability point,  has to eventually spin up.
This conclusion follows immediately from a  Lemma
formulated  by Friedman et al. (1988).
\footnote[2]{We note a misprint in this  Lemma text as printed in
Friedman et al. (1988), who use dots to denote a derivative
with respect to $\lambda$:  the dot over the right bracket in
$(\dot \Omega \dot J + \dot \mu \dot N\dot{)}\ne 0$ is missing.}
These authors considered a two-parameter family of uniformly rotating
stars with a one-parameter  EOS $P=P(\rho)$. In general, a continuous
sequence of rotating configurations can be labeled by a parameter
$\lambda$, so that along this sequence all stellar parameters are
functions of $\lambda$. In order to avoid confusion with time derivative
of a stellar quantity $Q$, we will denote a derivative of $Q$  with
respect to $\lambda$  by  ${Q}^\prime \equiv {\rm d}Q/{\rm d}\lambda$.
According to Friedman et al. Lemma,
the unstable region corresponds to the part of the sequence $J(f)$ for which
$ J^\prime   f^\prime >0$ (since along these sequences ${M}_{\rm B}^\prime
 =0$), or  equivalently, ${\rm d}J/{\rm d}f>0$, which means
that just before the instability is
reached  one has ${\rm d}J/{\rm d}f<0$, i.e., the  spin up by the angular
momentum loss.

Concluding, ${\rm BB}_{\rm ut}$ is a feature of any equation of
state. However the significance of this effect depends on the
stiffness of the matter. In contrast ${\rm BB}_{\rm st}$ exists
only when $\mb$ has a local minimum for fixed rotational
frequency.


\begin{figure}
\resizebox{\hsize}{!}{\includegraphics[angle=-90]{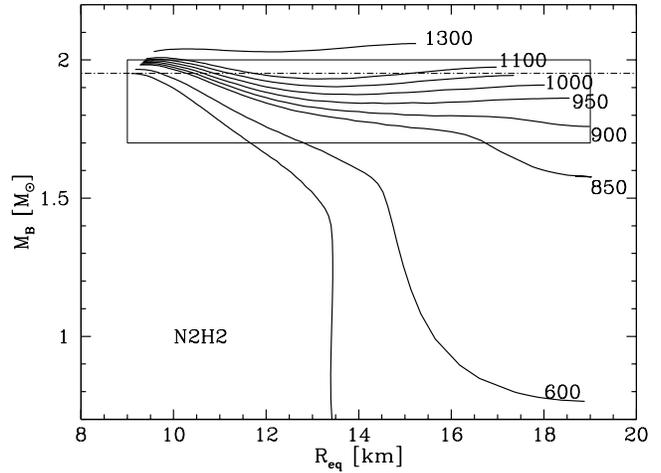}}
\caption{Model N2H2. Baryon-mass versus circumferential equatorial
radius for fixed frequency.
The minimum for fixed $\mb$ appears approximately at the same frequency
at which the decreasing part for large $R_{\rm eq}$
(close to Keplerian configurations) disappears,
i.e. this limiting case correspond to
the nearly flat curve ($\mb={\rm const}$) up to the Keplerian configuration.
There is no maximum for fixed frequency.}
\label{fig:mbrn2h2}
\end{figure}

\begin{figure}
\resizebox{\hsize}{!}{\includegraphics[angle=-90]{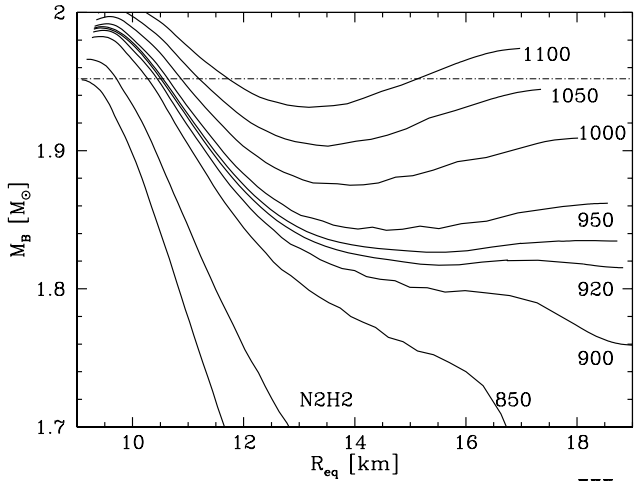}}
\caption{Model N2H2. The enlarged region
marked by the rectangular box  in Fig. \ref{fig:mbrn2h2}.
For $f\simeq 920$~Hz there exist region where the curve is nearly flat ($\mb$ does
not change) up to the Keplerian configurations.
For higher $f$ there exist a minimum of $\mb$.
The evolution of the isolated star which is losing its angular momentum
is represented by the
motion along horizontal line from right to the left (decreasing $J$).
For $\mb>1.82~\msol$ angular momentum loss leads to spin up.
For explanation of the ``way features'' of most curves see the text.
}
 \label{fig:mbrn2h2x}
\end{figure}

The impact of the hyperon EOS softening on the character and significance
of the BB phenomenon can be clearly appreciated  by
comparing Figs.\ \ref{fig:jfn2h1},\ \ref{fig:jfn1h1}
with  Fig.\ \ref{fig:jfn1} obtained for the (nucleonic)  N1 EOS without
hyperons. For the nucleonic EOS, BB is present only for
supramassive configuration and it is significant only for
relatively high masses (although strictly speaking it is
present for all masses larger than $M_{\rm B,max}^{\rm stat}$ --  for
masses $M_{\rm B}< 2.6~ {\rm M}_\odot$ the ${\rm BB}_{\rm ut}$
region is so
narrow  that it is not visible in Fig.\ \ref{fig:jfn1}), see
Cook et al. (1994) and Salgado et al. (1994). In contrast, in the
case of the EOS with hyperons, the BB is present for configurations close to the
Keplerian one for very broad range of masses.
The BB epoch for the nucleonic EOS terminates always by an instability.
\section{The EOS-dependence of the back bending phenomenon}
\label{sect:BB.EOS}
The specific form of the BB phenomenon depends on the degree of softening
of the EOS above the hyperon threshold. We considered a rather large set of
the hyperonic matter, including four models of Balberg \& Gal (1997)
described in Sect.\ \ref{sect:EOS},
and the five models of Glendenning (1985). Rapid rotation and spin-evolution
for  the N1H1 and N2H1 EOSs of Balberg \& Gal (1997)    and of
an EOS of
Glendenning (1985),  were previously studied, using different numerical method of
solving of the 2-D equations of hydrostatic equilibrium, by Balberg et al. (1999).

We obtain the most pronounced BB behavior for the N2H1 EOS, which
{\it was not studied}
by Balberg et al. (1999). In this model the nucleonic EOS is very stiff, and the
hyperon softening is not so strong as for the EOS with the H2 hyperonic sector.
Such features  are optimal for getting a strong BB effect for a large interval
of stellar baryon mass. In particular, we get  ${\rm BB}_{\rm st}$ for a relatively
large interval of normal rotating configurations with $M_{\rm B}=
(1.91-2.05)~{\rm M}_\odot$, which extends up to $2.17~{\rm M}_\odot$ if
supramassive rotating models are also included. At $M_{\rm B}\simeq 2~{\rm M}_\odot$
the star which starts its spin evolution at $f\simeq f_{\rm K}$ accelerates its
rotation by $\sim 50$~Hz  during the first epoch of its evolution when nearly half of
its initial angular momentum is lost ! Implications of such a substantial ``back
bending with stable termination'' for the pulsar timing interpretation will be
discussed in Sect.\ \ref{sect:DiscConcl}.

The  N1H1 EOS has significantly softer nucleon component, and more important
flattening of the $M_{\rm B}(R_{\rm eq})$ curve.
As for the N2H1 EOS, the  ${\rm BB_{st}}$ phenomenon  occurs not only for the
supramassive configurations but also for the normal ones,
with $M_{\rm B}<M_{\rm B,max}^{\rm stat}$
(Fig.\ \ref{fig:jfn1h1}). However, the range
of the baryon masses of the normal configuration for
which the ${\rm BB_{st}}$
 occurs is very narrow,  $\mblm=1.81~{\rm M}_\odot
 <M_{\rm B}<M_{\rm B,max}^{\rm stat}=1.823~
{\rm M}_\odot$. For such baryon masses ${\rm BB_{st}}$ takes place
at $f>\flm \simeq
850~$Hz.  The ${\rm BB_{\rm st}}$ itself for this EOS
is due to the high degree of flatness of the hyperon section
of the $M_{\rm B}(R_{\rm eq})$
plot. However, due to this flatness the values of $\mblm$ and
$M_{\rm B,max}^{\rm stat}$ are also very close to each other.
 All in all, even including the supramassive  configurations, the range of
baryon masses for which the ${\rm BB_{st}}$ occurs is not large,
$1.81~{\rm M}_\odot<M_{\rm B}<1.9~{\rm M}_\odot$.

The situation changes if we consider the N2H2 EOS (see Sect.\ \ref{sect:EOS}),
where the hyperon softening is stronger.
Let us repeat the analysis applied previously to N1H1 EOS
in Sect.\ \ref{sect:MR.minima}. The $M_{\rm B}(R_{\rm eq})$ curves at several
values of $f$ are shown in Fig.\ \ref{fig:mbrn2h2}. In contrast to the case
of the N1H1 EOS,  pictured in Fig.\ \ref{fig:mbrn1h1} there is
 no curve with a  visible  second (large-radius) maximum. A zoom of the relevant
 baryon-mass -- equatorial-radius rectangle is displayed
 in Fig.\ \ref{fig:mbrn2h2x}. A  marginal second maximum
 appears at $f\simeq 920$~Hz. However, this second maximum is not a
 robust one. It can be detected only if the precision of the calculation
 is sufficiently high. This is to be contrasted with the N1H1 case, where the
 second maxima on the $M_{\rm B}-R_{\rm eq}$ curves  are well pronounced,
 see Fig.\ \ref{fig:mbrn1h1x}. Let us notice that
 N2H2 is the EOS for which the BB phenomenon for normal configurations was detected
 by Balberg et al. (1999).


\newcommand {\st}{\rightarrow}
\begin{table}[t]
\caption{
Main parameters relevant for  the spin evolution of isolated  neutron stars with
hyperons.
The labels ``infx'' mark  the configurations for which the curve
$\mb(R_{\rm eq})_{f}$ has a
point of inflexion. For
$f_{\rm 1,infx}<f<f_{\rm 2,infx}$ (corresponding
masses $M_{\rm B}^{\rm (1,infx)}$ and
$M_{\rm B}^{\rm (2,infx)}$) ${\rm BB_{st}}$ exists.
For $M_{\rm B}>M_{\rm B,max}^{\rm stat}$ we have  also ${\rm BB_{ut}}$,
which is the only BB phenomenon  for
$\mb > M_{\rm B}^{\rm (2,infx)}$. For $f>f_{\rm K,flat}$
and  $M_{\rm B}>M_{\rm B}^{\rm (K,flat)}$, a star which is initially in the mass-shedding
(Keplerian) state, spins-up   as it looses angular momentum. The characteristic
S-shape of $J(f)$ is then impossible.}
\label{tab:bbhyp}
\begin{center}
\begin{tabular}{ccccc}
\hline\hline
Parameter   & N1H1 & N2H1 & N1H2  &N2H2 \\
\hline
 $M_{\rm B,max}^{\rm stat}[\msol]$& 1.83 & 2.05 & 1.73 & 1.95\\
\hline
 $f_{\rm 1,infx}$ [Hz]& 860 & 880 & 880 & 910\\
 $M_{\rm B}^{\rm (1,infx)}[\msol]$& 1.808 & 1.91 & 1.73 & 1.81\\
\hline
 $f_{\rm K,flat}$ [Hz]     & 1030 & 970 & 1050 & 930\\
 $M_{\rm B}^{\rm (K,flat)} [\msol]$& 1.96 & 2.01 & 1.87  & 1.82\\
\hline
 $f_{\rm 2,infx}$ [Hz]& 1100 & 1320 & 1080 & 1380\\
 $M_{\rm B}^{\rm (2,infx)} [\msol]$& 1.9 & 2.17 & 1.79 & 2.06\\
\hline
\hline
\end{tabular}
\end{center}
\end{table}


The main features of the dependence of the back bending phenomenon
on the  EOS  can be summarized on the basis of the Table
\ref{tab:bbhyp} containing relevant  parameters  for considered
EOSs. Let us start with four characteristic baryon masses $M_{\rm
B}^{\rm (1,infx)}$, $M_{\rm B}^{\rm (2,infx)}$, $M_{\rm B}^{\rm
(K,flat)}$, $M_{\rm B,max}^{\rm stat}$. Within this set, the
ordering according to the baryon-mass value turns out to depend
{\it only} on the pure nucleon EOS. The sequence  for N1H1 and
N1H2  EOSs is: $M_{\rm B}^{\rm (1,infx)}<M_{\rm B,max}^{\rm stat}
<M_{\rm B}^{\rm (2,infx)}<M_{\rm B}^{\rm (K,flat)}$ whereas for
the N2H1 and N2H2 EOSs  we have $M_{\rm B}^{\rm (1,infx)}<M_{\rm
B}^{\rm (K,flat)} <M_{\rm B,max}^{\rm stat}<M_{\rm B}^{\rm
(2,infx)}$. As can be seen the main difference is a  rather low
value of $M_{\rm B}^{\rm (K,flat)}$ for the N1H1 and N1H2 models.
In the case of the N2 model of the nucleon component,  rotation
changes the properties of the star close to the Keplerian
frequency much more effectively
 than for the N1 one.

The nucleon  N2 EOS is stiffer than the N1 one. As a result,
 the stars with nucleon envelopes based on the N2 model  are more extended
 than the  N1 ones, having larger radius at the same mass.
 On the other hand,  the density profiles in the nucleon envelopes
of the  N2H (H=H1,H2)  stars  have  smaller radial gradients,  and
therefore they  play more  important role during rotation. For example,
the N2 envelopes have significantly  larger mass than the N1 ones.
Consequently, a  smaller rotational frequency ($f_{\rm K,flat}= 930, 970$ Hz
for H2, H1
softening, respectively),
than in the case of the N1 model,
is needed
to make the effects of rotation so important that the maximum of $M$ close
to Keplerian frequency disappears.
This effect can be easily seen in Table\ \ref{tab:bbhyp} where not only the
N1H  frequencies $f_{\rm K,flat}$ are significantly larger than the
N2H ones, but also the mass $M_{\rm B}^{\rm (K,flat)}$ is significantly
larger. Not only it exceeds the maximum mass for the nonrotating stars,
 $M_{\rm B,max}^{\rm stat}$, but it is even larger than
 $M_{\rm B}^{\rm (2,infx)}$. In this case the
$M_{\rm B}^{\rm (2,infx)}$ determines the mass above which only
spin up by angular momentum loss is possible.

As the N1 EOS is softer than the N2 one,   the  additional softening by
hyperons leads  almost immediately
to the maximum mass of nonrotating configurations; the hyperon segment
beyond the ``hyperon knee'' is very flat.
The back bending phenomenon ${\rm BB_{st}}$  is present between the
two  inflexion points,  and in general is not connected with the maximum mass
of nonrotating stars.

 The $M_{\rm B}(R_{\rm eq})$ curves displayed in Fig.\ \ref{fig:mbrn2h2x}
deserve  an additional comment,  referring to the precision of the
2-D calculations.  Nearly all curves for $f=850-1100$~Hz  (except
for two curves  for  $f\simeq 920$~Hz) were obtained in numerical
calculations in which the innermost  zone  boundary {\it is not
adjusted} to the surface of the hyperon threshold.  Consequently,
these  curves exhibit ``wavelets'' which result from an
insufficient precision of the numerical calculations. In contrast,
the two curves  for $f\simeq 920$~Hz  are calculated with the
innermost zone boundary at the  hyperon-softening threshold, which
enables a much higher precision.

The division of the stellar interior  into two zones is
particularly effective in the case of a strong change in the  EOS
 $P(\rho)$ at some density.  In our case we have a rather  stiff EOS
below the threshold density for the hyperon appearance, and a
 soft EOS for dense matter with hyperons. Therefore, we put the
inner zone boundary at the threshold of hyperon appearance, where
the adiabatic index of the EOS suffers a significant drop. This allows
us  for very accurate calculation by the spectral method also in the
region close to the hyperon threshold.  Although in our case
we do not encounter a density jump at the zone boundary, our method
can be also used for an EOS with a density jump due to a first-order
phase transition.

\section{Discussion and conclusion}
\label{sect:DiscConcl}
The presence of hyperons in neutron-star cores  can strongly affect the spin
evolution of a solitary neutron star (e.g., an isolated pulsar). As we have
shown, such a neutron star can live  a long epoch  of ``spin down by the
angular momentum loss'', and this could occur for a broad range of baryon mass
of neutron star. The epoch with
$\dot{P}<0$ despite $\dot{J}<0$
 could terminate
by an instability or by a stable continuous transition to a ``standard''
spin-down epoch.

We paid particular attention to the $(\dot{P}<0$, $\dot{J}<0,M_{\rm B}=const)$
epoch with a stable termination. It is  represented by an S-shaped segment
of the spin-evolution track  in the  $J-f$ and $I-f$ planes,  and was baptized
``back bending phenomenon'' in the previous literature. Various regimes
of the spin evolution were shown to be correlated with the behaviour of the
$M_{\rm B}(R_{\rm eq})_f$ curves at fixed $f$. In particular, we pointed out
importance of the location of the inflection points of the
$M_{\rm B}(R_{\rm eq})$ curves for the existence of the back bending
 phenomenon.  We were also inclined to leave the name of back bending
only to the evolution-track segments with a stable termination
(${\rm BB_{st}}$ in our terminology).

Epochs with back bending for normal rotating configurations
were found for two of the four EOS of Balberg \& Gal (1997). On the
other hand, we found that the back bending phenomenon for normal sequences was absent
for five hyperonic EOSs of Glendenning (1985). This illustrates the
uncertainties in the hyperonic EOS, stemming for a high degree of
ignorance concerning the nucleon-hyperon and particularly
hyperon-hyperon interactions in dense matter.

Throughout this paper we stressed that the back bending  in the $I-f$
plane, considered up to now in the literature, should be accompanied
by stability analysis; only stable back-bending configurations are
astrophysically relevant. We performed such a stability analysis, and
we found that very often  dominant back-bending segments of the
$I(f)_{\rm M_{\rm B}}$  are {\it unstable} with respect to the axisymmetric
perturbations, and therefore do not exist in the Universe.

\vskip 3mm

As it has  been mentioned by Spyrou \& Stergioulas (2002), in such
kind of calculations it is extremely important to assure the
thermodynamical consistency of the EOS (the first law of
thermodynamics has to be strictly fulfilled). It is well known
that a rough treatment of this condition can lead to an inaccurate
determination of the maximum mass and  stability conditions  (for
example the configurations corresponding to the maximum of  $M$
and $M_{\rm B}$ do not coincide - this is an evidence for  a lack
of the thermodynamical consistency in the EOS).

The precision of the code is also very important and in our case
the proper division of the star into two computational domains
(at the threshold for
appearance of hyperons) allows us to obtain high precision results without a
large increase of the number of grid points.

\vskip 3mm

An isolated pulsar, born in a SN II explosion, could have an interesting
and nonstandard past due to a hyperon softening of the EOS.
As we showed, such a pulsar could lose some half  of its initial
angular momentum without changing much its  rotation period. Therefore,
if one observes a rapid pulsar with a characteristic age $\tau_{\rm PSR}$
significantly longer than the age of a supernova remnant where this pulsar
is born, $\tau_{\rm SNR}$, this might be due to some back bending episode,
resulting from a  (hyperon? phase-transition?) softening of the EOS of
its core. Such a possibility of explaining a seeming contradiction
between  $\tau_{\rm SNR}$ and $\tau_{\rm PSR}$ has been already noticed,
in the context  of a  mixed-phase EOS softening, by \cite{ss02}.
Clearly,
the observational pulsar-timing signatures  of the EOS softening due
to the hyperon or phase-transition softening of the EOS deserve further
studies, and we are planning to continue such studies using our high-precision
2-D code.

Another interesting consequence of the hyperon softening of the EOS could
be a ``period clustering'' of rotating neutron stars  powered by accretion
in the long living low-mass X-ray binaries. This problem is now being
studied.

\acknowledgements{We are very grateful to Nick Stergioulas for helpful
correspondence on the spin evolution of solitary neutron stars, and
for making possible comparison of numerical results obtained by our two
groups. We express our gratitude to John L. Friedman, who during a car
ride from Orsay to Meudon gave us a precious advice concerning the
axisymmetric instabilities in rotating neutron stars. Last but not
least, we  thank Brandon Carter whose careful driving made such a
fruitful discussion possible.\par
This work was partially supported by KBN grants 5P03D.020.20
and 2P03D.019.24.}



\begin{thebibliography}{} 

\bibitem[Ambarts60]{Ambarts60}
Ambartsumyan, V. A \& Saakyan G.S, 1960, Astron. Zhur., 37, 193 (English translation
in Sov. Astron.-AJ, 4, 187)

\bibitem[BalbGal1997]{BalbGal1997}
Balberg, S. \& Gal, A., 1997, Nucl.Phys. A,  625, 435

\bibitem[Balberg et al.(1999)]{blc99}
Balberg, S., Lichtenstadt, I., \& Cook, G.P, 1999, ApJ, 121, 515

\bibitem[Bonazzola et al. (1998)]{BonazGM98}
Bonazzola, S., Gourgoulhon, E., \& Marck, J.-A., 1998,
Phys. Rev. D, 58, 104020

\bibitem[Bonazzola et al. (1993)]{BonazGSM93}
Bonazzola, S., Gourgoulhon, E., Salgado, M., \& Marck, J.-A.,
1993, A\&A, 278, 421

\bibitem[Cameron1959]{Cameron1959}
Cameron, A.G.W, 1959, ApJ, 129, 676

\bibitem[Cheng et al.(2002)]{cyz02}
Cheng, K. S., Yuan, Y. F. \& Zhang, J. L., 2002, ApJ, 564, 909

\bibitem[Chubarian2000]{Chubarian2000}
Chubarian, E., Grigorian, H., Poghosyan, G., Blaschke, D.,
2000, A\&A, 357, 968

\bibitem[CookST1994]{CookST1994}
Cook, G.B., Shapiro, S.L., Teukolsky, S.A., 1994, ApJ,
423, L117

\bibitem[Friedman et al. 1988]{Fried1988}
Friedman, J.L., Ipser, J.R, Sorkin, R.D., 1988,
ApJ, 325, 722

\bibitem[Glendenning et al.(1997)]{gpw97}
Glendenning, N. K., Pei, S. \& Weber, F., 1997, Phys. Rev. Lett.,
79, 1603

\bibitem[Glendenning2001]{Glen01}
Glendenning, N.K., 2001, J. Phys. G: Nucl. Part. Phys., 28, 2023

\bibitem[GlendWeber2002]{gw02}
Glendenning, N.K. \& Weber, F., 2002, AIP Conference Proceedings, 610, 470

\bibitem[Gourgoulhon et al. (1999)]{GourgHLPBM99}
Gourgoulhon, E., Haensel, P., Livine, R., Paluch, E., Bonazzola, S.,
\& Marck, J.-A., 1999, A\&A, 349, 851

\bibitem[Haensel2003]{Haensel03}
Haensel, P., 2003, in: Final Stages of Stellar Evolution, eds. J.-M.
Hameury \& C. Motch, EAS Publications Series vol. 7 (EDP Sciences)
p. 249

\bibitem[HaenselLevenfish2002]{hl02}
Haensel, P., Levenfish, K.P., \& Yakovlev, D.G., 2002,
A\&A 394, 213


\bibitem[Nozawa et al. (1998)]{NozawSGE98}
Nozawa, T., Stergioulas, N., Gourgoulhon, E., \& Eriguchi, Y., 1998,
A\&AS 132, 431

\bibitem[Ring \& Schuck (1980)]{Ring80}
Ring, P. \& Schuck, P., 1980, The Nuclear Many Body Problem, Springer, Berlin

\bibitem[Salgado94]{1994}
Salgado, M., Bonazzola, S., Gourgoulhon, E., Haensel, P., 1994, A\&A, 291, 155

\bibitem[Salpeter1960]{salp60}
Salpeter, E.E., 1960, Ann.Phys., 11, 393


\bibitem[Spyrou \& Stergioulas (2002)]{ss02}
 Spyrou, N.K. \& Stergioulas, N., 2002, A\&A, 395, 151

\bibitem[Stergioulas (2003)]{ster02}
 Stergioulas, N., 2003, Living Rev. Relativity, 6, 3,
http://www.livingreviews.org/lrr-2003-3

\bibitem[Swesty (1996)]{Swesty96}
Swesty, F.D., 1996,  J. Comp. Phys., 127, 118

\end{thebibliography}
\end{document}